\documentclass[conference]{IEEEtran} 
\ifCLASSINFOpdf
\else
\fi
\usepackage{cite}
\usepackage{amsmath}
\usepackage{amssymb}
\usepackage{amsfonts}
\usepackage{psfrag}
\usepackage{epsfig} 
\usepackage{algorithm}
\usepackage{algorithmic}
\usepackage{amsthm}

\newcommand{\B}[1]{\boldsymbol{#1}}
\begin{document}
%
\title{mmWave Massive MIMO with Simple RF and Appropriate DSP} 


\author{\IEEEauthorblockN{Amine Mezghani}
\IEEEauthorblockA{Wireless Networking and Communications Group\\
Department of ECE, University of Texas at Austin\\
Austin, TX 78712, USA\\ 
Email: amine.mezghani@utexas.edu} 
\and
\IEEEauthorblockN{A. Lee Swindlehurst}
\IEEEauthorblockA{Center for Pervasive Communications and Computing\\
Department of EECS, University of California, Irvine\\
Irvine, CA 92697, USA\\
Email: swindle@uci.edu} }

\maketitle

\begin{abstract}
There is considerable interest in the combined use of millimeter-wave (mmwave) frequencies and arrays of massive numbers of antennas (massive MIMO) for next-generation wireless communications systems. A symbiotic relationship exists between these two factors: mmwave frequencies allow for densely packed antenna arrays, and hence massive MIMO can be achieved with a small form factor;  low per-antenna SNR and shadowing can be overcome with a large array gain; steering narrow beams or nulls with a large array is a good match for the line-of-sight (LOS) or near-LOS mmwave propagation environments, etc.. However, the cost and power consumption for standard implementations of massive MIMO arrays at mmwave frequencies is a significant drawback to rapid adoption and deployment. In this paper, we examine a number of possible approaches to reduce cost and power at both the basestation and user terminal, making up for it with signal processing and additional (cheap) antennas. These approaches include low-resolution Analog-to-Digital Converters (ADCs), wireless local oscillator distribution networks, spatial multiplexing and  multi-streaming instead of higher-order modulation etc.. We will examine the potential of these approaches in making mmwave massive MIMO a reality and discuss the requirements in terms of digital signal processing (DSP).
\end{abstract}


%
\IEEEpeerreviewmaketitle

\section{Introduction}  
\begin{figure}[h]
\psfrag{a}[c][c]{(a)}
\psfrag{b}[c][c]{(b)}
\centerline{\includegraphics[width=5cm]{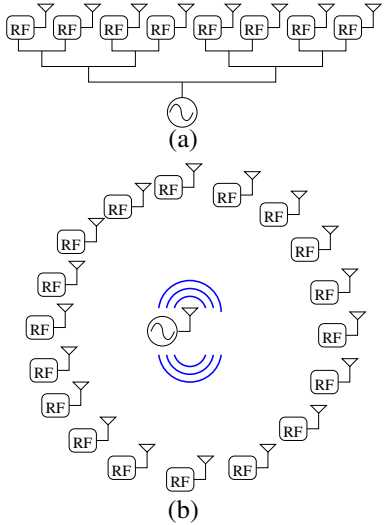}}
\caption{LO distribution network: (a) Standard implementation (b) Wireless synchronization.}
\label{wireless_syn}
\end{figure}

 The deployment of large numbers of possibly on-chip integrated or distributed antennas, known as massive Multiple Input Multiple Output (MIMO) systems, the access to more bandwidth through the use of mmwave frequencies, and the use of low-cost wired and wireless optical links are all considered as key enablers to meet the impending demand for gigabit per second wireless data.  However the implementation of these large and sophisticated wireless systems will lead to a significant increase in  cost, complexity and power dissipation. In particular, synchronization, local oscillator (LO) generation and distribution is very challenging and the need to  feed  each antenna with the LO signal for the demodulation process as shown in Fig.~\ref{wireless_syn}(a) is a very critical issue. Due to the  losses of electrical wires that increase with trace length and frequency, we propose and analyze, as alternative, the concept of wireless LO distribution for large MIMO receivers, following the approach in \cite{Mezghani_letter}.   The basic idea is to deploy a ``dummy" radiating antenna placed at a distance (several wave lengths) away from the intended receiving antenna array to radiate, and thus wirelessly broadcast,  a very weak carrier signal for synchronization and direct detection, as depicted in Fig.~\ref{wireless_syn}(b). Then, the receiving antenna elements can apply simple additive mixing, e.g. direct detection with a diode,  and have almost perfect access to the in-phase and quadrature signals of the users. 
A major advantage of wireless is the fact that the LO power attenuation follows just the inverse square law of free space propagation with respect to the array dimension, while it suffers from an exponential attenuation with respect to the trace length in wired synchronization due to the skin-effect. Due to regulatory restrictions, however, the radiated LO power has to be kept at a very low level.  In this paper we show that the achievable rate of this RF architecture with weak wireless LO power  and simple direct detection antennas can still approach the ideal performance by using asymmetric low order bandpass filtering.  This solution also offers the advantage of integrating the complete millimeter-wave demodulation circuit into the antenna for better sensitivity, and enabling the use of ``cheap" active antennas.  \\

Furthermore, the idea of wirelessly synchronized antennas can be combined with the utilization of low-resolution, for instance,  one-bit ADCs that  essentially possess a unique circuit implementation and might therefore qualify as a fundamental research area in modern communication theory. In fact, the analysis of the quantization process has gained a lot of attention in academic research  \cite{mezghaniisit2010,mezghaniisit2007,mezghaniisit2009,Hea14, heath2015, Mollen, Jacobsson_2017}. The proposed front-end structure with combined wireless synchronization and one-bit ADCs can be significantly simplified by employing such low performance devices. Since just one-bit ADCs is used, no further amplification is needed after the power detection. 
This avoids the issues of other analog implementations (calibration, power, chip-area, aging, etc.) and leads to a  very cost and energy efficient front-end implementation with direct high speed speed digital output, which we referred to as a direct digitization one-bit antenna.  Surprisingly, at low SNR, prior to processing like beamforming,  the loss due to one-bit quantization is approximately equal to only $\pi/2$ (1.96dB) in conventional MIMO systems regardless of the type of available channel state information \cite{mezghaniisit2007,mezghaniisit2009,mezghani_2012_isit}. We analyze in this paper the validity of this result in the context of massive MIMO and deduce some implications in terms of DSP requirements.

\emph{Notation:}
Vectors and matrices are denoted by lower and upper case italic bold letters.  The operators $(\bullet)^\mathrm {T}$, $(\bullet)^\mathrm {H}$, $\textrm{tr}(\bullet)$ and $(\bullet)^*$ stand for transpose, Hermitian (conjugate transpose), trace, and complex conjugate, respectively.  The terms $\B{1}_M$ and ${\bf I}_M$ represent the all ones vector and the identity matrix of size $M$, respectively. The vector $\boldsymbol{x}_i$ denotes the $i$-th column of a  matrix $\B{X}$ and $\left[\B{X}\right]_{i,j}$ or $x_{i,j}$ denotes the ($i$th, $j$th) element, while $x_i$ is the $i$-th element of the vector $\B{x}$.  We represent the Hadamard (element-wise in each real dimension) and $n$-th power of vectors and matrices by the operators "$\B{A}\circ \B{B}$" and "$\B{A}^{\circ n}$", respectively, i.e., $[\B{A}\circ \B{B}]_{i,j}={\rm Re}[\B{A}]_{i,j}  {\rm Re}[\B{B}]_{i,j} + {\rm j} {\rm Im}[\B{A}]_{i,j}  {\rm Im}[\B{B}]_{i,j}$. Additionally, $\textrm{diag}(\boldsymbol{B})$ denotes a diagonal matrix containing only the diagonal elements of $\boldsymbol{B}$ and $\textrm{nondiag}(\boldsymbol{B})=\boldsymbol{B}-\textrm{diag}(\boldsymbol{B})$. Further, we define $\B{C}_x={\rm E}[\B{x}\B{x}^{\rm H}] - {\rm E}[\B{x}]{\rm E}[\B{x}^{\rm H}]$ as the covariance matrix of $\B{x}$ and $\B{C}_{xy}$ as $\mathrm {E}[\B{x}\B{y}^{\rm H}]$. Finally, $\mathcal{O}(\cdot )$ and $o(\cdot )$ represent the Bachmann-Landau notation for asymptotic behaviors.

\section{Analysis of the Wireless Synchronization Architecture}
\label{sec:system}

In the proposed wireless LO distribution system with additive mixing, the receiver needs  an appropriate  way to lift the weak carrier to a higher power level  compared to the data signal such that the mixing at the diode occurs almost with negligible intermodulation products leading to an intact  signal in the  baseband. This enables approaching ideal performance with a low power penalty. However, to achieve this goal,  a popular approach widely used in the literature and in practice is to include an additional separate processing path to recover a replica of the residual carrier, typically using active circuits for phase injection-locking \cite{Malyon}. Then, the filtered  signal and the reconstructed carrier are recombined before being applied to the diode (or a mixer). The resulting complexity is, however, still substantial for massive MIMO. In contrast, we show next that, interestingly, even a simple purely passive bandpass filter $G(f)$ of low order (smooth $1/f$ passband characteristic) is potentially sufficient for joint processing of the carrier and data signal and approaching ideal performance without isolating the LO signal.  \\

\begin{figure}[tb]
\centering
\psfrag{y}[c][c]{$ y(t)$}
\psfrag{z}[c][c]{$z(t)$}
\psfrag{R}[c][c]{$G(f)$}
\psfrag{|.|}[c][c]{$|\bullet|^2$}
\psfrag{x}[r][r]{$d_k  $}
\psfrag{s}[c][c]{$x$}
\psfrag{sL}[c][c]{$\varepsilon x_{\rm LO}$}
\psfrag{n}[c][c]{$z$}
\psfrag{r}[c][c]{~~~~~~~$r_{\rm DD}[n]$}
\psfrag{2B}[c][c]{$2B$}
\psfrag{B}[c][c]{\tiny $\frac{B}{2}$}
\psfrag{-B}[c][c]{\tiny $\!\! -\frac{B}{2}$}
\hspace{-1mm}\includegraphics[width=0.43\textwidth]{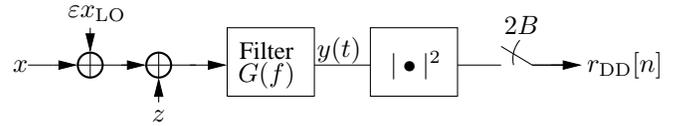}
\vspace{-2mm}
\caption{Baseband representation of additive mixing at each antenna.\vspace{-0.5cm}}
\label{baseband}
\end{figure}

Since a real valued bandpass signal and its complex baseband envelope representation have the same magnitude/absolute value, we represent all the signals in the baseband for simplicity. The baseband model of each direct detection antenna is shown in Fig.~\ref{baseband}. Assuming for simplicity the LOS case and ignoring the array phase shift, the individual antenna signal is described in the frequency domain as
\begin{equation}
 Y(f) = G(f) \left(X(f) + X_{\rm LO}(f)  +Z(f)\right),    
\end{equation}
with the information signal $X(f) $ having power spectral density $\Phi_x(f)$ limited to  bandwidth $B$ and the noise $Z(f)$ having the  power spectral density $\Phi_n(f)=N_0$, while $G(f)$ is the transfer function of the receive bandpass filter (BP) in the baseband representation. The additive mixing with the LO signal is obtained by taking the instantaneous power (squared-magnitude) of $y(t)$ and sampling it at $2B $ Hertz:
\begin{equation} 
r_{DD}[n]=\left|y\left(n\frac{1}{2B}\right)\right|^2, 
\end{equation}
where $n$ is the discrete time index and $2B$ is the Nyquist sampling frequency.  In a coherent receiver the filter $G(f)$ plays the role of a band limiting anti-aliasing filter where the shape in the passband is irrelevant from a  theoretical point of view as long as it is reversible. However as shown later, in the additive mixing case, the shape of this filter is crucial.

 Further, let us assume  that $x(t)$ consists of a complex Gaussian process with a rectangular spectral density that is band-limited to the two-sided bandwidth $B$. In addition, in the proposed wireless synchronization architecture,  a monotone carrier signal is superimposed at the boundary of (or outside) the information signal band by means of a ``dummy" antenna. The PSD of the noiseless received signal reads then as 
\begin{equation}
 \Phi_{x+x_{\rm LO}}(f)= \frac{\rho}{B} \cdot {\rm rect} (f/B) +  \varepsilon P_{\rm LO} \cdot \delta(f-\frac{B}{2}),
 \label{phix} 
\end{equation}  
where $0<\varepsilon<1$ is the portion of the radiated carrier tone power $P_{\rm LO}$ captured on each antenna and  the rectangular function ${\rm rect}(f)$ equals 1 for $|f|\leq 1/2$ and is 0 elsewhere. Due the strict regulatory restrictions in terms of radiated power spectral density (especially for receivers),  the power $P_{\rm LO}$ of the spectrally peaky LO signal has to be very low ($\leq 1\mu$W). Additionally, one could alternatively aim at using an LO frequency outside the authorized band to downconvert the signal to an intermediate frequency with more restricted emission limits but for easier filtering and better performance than the tight choice in (\ref{phix}). Together with the fact that the attenuation factor $\varepsilon$ decreases with the number of antennas, this does not ensure the dominance of the LO signal compared to the noise and information signal and consequently the use of additive mixing at each antenna is not obvious. Therefore, a careful analysis of this concept is required. 

\begin{figure}
\centering
\psfrag{R}[c][c]{$R$}
\psfrag{L}[c][c]{$L$}
\psfrag{C}[c][c]{$C$}
\psfrag{R1}[c][c]{$R_1$}
\psfrag{L1}[c][c]{$L_1$}
\psfrag{C1}[c][c]{$C_1$}
\psfrag{BP}[c][c]{$\quad\quad\quad\quad\quad~ G_{\rm \!BP\!}(f)\!=\!G(|f|\!-\! f_0)$}
\hspace{-1mm}\includegraphics[width=0.4\textwidth]{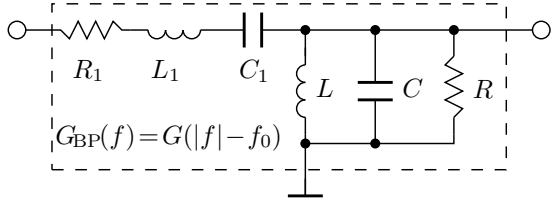}
\vspace{-2mm}
\caption{RLC-ladder circuit for the bandpass filter.\vspace{-0.5cm}}
\label{rlc}
\end{figure}

\begin{figure}
\centering
\psfrag{G(f)}[c][c]{$|G(f)|^2$}
\psfrag{f}[c][c]{$f/B$}
\hspace{-1mm}\includegraphics[width=0.4\textwidth]{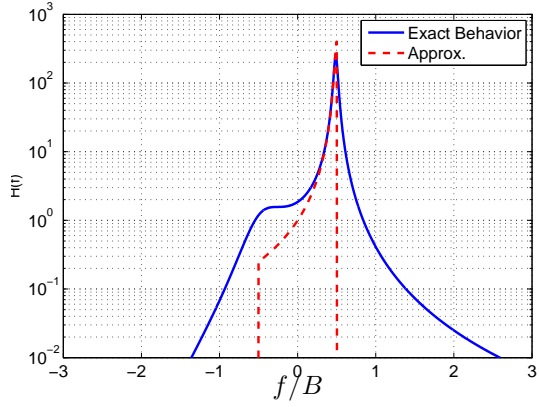}
\vspace{-2mm}
\caption{Example of asymmetric baseband transfer function of the RLC BP in Fig.~\ref{rlc} {  with $R_1^{-1}\sqrt{L_1/C_1}=80$, $R\sqrt{C/L}=4$, $(2\pi \sqrt{L_1C_1})^{-1}\!-\!(2\pi \sqrt{\!LC})^{-1}\!=\!B$}, and its approximation in (\ref{approx_transfer}) for  $\sigma\!=\!\frac{R\sqrt{C/L}}{R_1^{-1}\!\sqrt{\!L_1\!/C_1}}\!=\!0.05$.\vspace{-0.3cm}}
\label{transferfunction}
\end{figure}

To cope with this issue and enhance the LO signal prior to mixing, we utilize the following asymmetric receive filter transfer function in the baseband representation with $1/f$ first order roll-off rate in the passband:
\begin{equation}
G(f)= \left\{
\begin{array}{ll}
	0 &  \textrm{for } |f|    > \frac{B}{2}  \\
 \displaystyle	\frac{1}{ {\rm j}(2f/B-1) +  \sigma  }  &  \textrm{for }  |f|  \leq \frac{B}{2},
\end{array}
   \right. 
   \label{approx_transfer}
\end{equation}
with a parameter $\sigma>0$ called the dissipation factor or inverse Q-factor. This filter corresponds in the passband to a simple second order RLC filter with resonant frequency around $f_0\!+\!\frac{B}{2}$ ($f_0$ is the center frequency) multiplied with an ideal bandpass filter of bandwidth $B$. A tentative approximate implementation of the desired frequency behavior using RLC-ladder circuits is illustrated in Fig.~\ref{rlc} and \ref{transferfunction} and consists of a cascade of RLC series and parallel circuits with slightly different resonant frequencies, i.e. $|\frac{1}{2\pi \sqrt{LC}}-\frac{1}{2\pi \sqrt{L_1C_1}}|\!\propto \!B$, while one of the resonant circuits has an increasingly higher Q-factor { $1/\sigma$} (lower resistive loss) when $\varepsilon$ decreases. The stopband behavior of the filter is ignored in the model (\ref{approx_transfer}) (assumed to be ideal) since it only affects the noise reduction and our focus is instead on the self-interference issue.  

In general the achievable rate $R_{\rm DD}$ with wireless LO mixing, which is upper bounded by the ideal Shannon limit $\log_2(1+N\rho/(BN_0))$ with $N$ antennas,  is not trivial to determine, we will derive a  capacity lower bound assuming a Gaussian input, and show that this lower bound can arbitrarily approach the upper bound under { the assumption} of certain asymptotic behavior for $G(f)$ in (\ref{approx_transfer}).
At the output $y(t)$ of this analog receive filter, we get the following PSD for $y(t)$ in each antenna 

\begin{equation}
    \Phi_{y}(f)  =  |G(f)|^2 \cdot \left( \Phi_{x+x_{\rm LO}}(f) + N_0   \right).
\label{phiy}    
\end{equation}
Next, the Fourier transform of the energy-detected signal $|y(t)|^2$ reads as the following convolution 
\begin{equation}
  \mathcal{F}\{|y(t)|^2\}=Y(f) \otimes  Y(-f)^*.
\end{equation}
{  Given that $Y(f)$ is uncorrelated across frequency, the PSD of  $|y(t)|^2$} is obtained as 

\begin{equation}
\begin{aligned}
& \Phi_{|y|^2}(f) = \lim\limits_{T \!\rightarrow \infty} \!\frac{1}{T}{\rm E} [|\mathcal{F}\{|y(t)|^2\}|^2]    \\
                 &= \lim\limits_{T \!\rightarrow \infty}\!\frac{1}{T} {\rm E} \! \left[\int \! Y(f') Y(f'\!-\!f)^* {\rm d}f' \!\int \! Y(f'')^* Y(f''\!-\!f)^* {\rm d}f'' \! \right] \\
                 &= \lim\limits_{T \!\rightarrow \infty}\!\frac{1}{T} {\rm E} \!\left[\int\!\!\int Y(f') Y(f'\!-\!f)^*  Y(f'')^* Y(f''\!-\!f)^* {\rm d}f'{\rm d}f'' \! \right] \\
                 &\stackrel{f\neq 0}{=} \lim\limits_{T \!\rightarrow \infty}\frac{1}{T^2}  {\rm E} \left[\int |Y(f')|^2 |Y(f'-f)|^2 {\rm d}f'  \right]  \\
                 &{=}\lim\limits_{T \!\rightarrow \infty}\frac{1}{T^2} \int {\rm E}[|Y(f')|^2] {\rm E}[|Y(f'-f)|^2] {\rm d}f'   \\
                 &{=} \int \Phi_{y}(f') \Phi_{y}(f'-f) {\rm d}f'   {=}  \Phi_{y}(f)	\otimes  \Phi_{y}(-f),   
 \end{aligned}
\end{equation}
where the last four steps hold for $f\neq 0$ and follow {  by expressing integrals as Riemann sums (${\rm d}f''\!=\!\frac{1}{T}$)}. In summary, 
\begin{equation}
\Phi_{|y|^2}(f) = \left\{
\begin{array}{ll}
\displaystyle
	\delta(f) \left(\int_{-\frac{B}{2}}^{-\frac{B}{2}} \Phi_{y}(f) {\rm d} f\right)^2 &  \textrm{for } f=0  \\
	\Phi_{y}(f)	\otimes  \Phi_{y}(-f) &  \textrm{for } f \neq 0.
\end{array}
   \right. 
\end{equation}

Since the DC component at the output of the energy detector is irrelevant to the information rate\footnote{The DC component is removed by a second filter after diode mixing in order to fully exploit the ADC dynamic range.}, we restrict the following analysis to the case of $f \neq 0$.  Using (\ref{phix}) and (\ref{phiy}), we obtain  for $f \neq 0$
\begin{equation}
 \begin{aligned}
   \Phi_{|y|^2}(f) \stackrel{f\neq 0}{=} & \varepsilon P_{\rm LO} \left|G(\frac{B}{2})\right|^2   \left|G(\frac{B}{2}-|f|)\right|^2 \!\!\left(N_0 + \frac{\rho}{B}\right) + \\
   &~~ |G(f)|^2 \otimes |G(-f)|^2 \left(N_0 +  \frac{\rho}{B}\right)^2    \\   
   \stackrel{f\neq 0}{=} &  \frac{\varepsilon P_{\rm LO}}{\sigma^{2}}    \left|G(\frac{B}{2}-|f|)\right|^2 \left(N_0 +  \frac{\rho}{B}\right) + \\
   &~~ |G(f)|^2 \otimes |G(-f)|^2 \left(N_0 +  \frac{\rho}{B}\right)^2.    
    \end{aligned}         
    \label{recy2}                 
\end{equation}
We notice that the first term in (\ref{recy2}) corresponds to a scaled version of the undistorted { received} signal, while the second term includes the non-Gaussian signal-to-signal intermodulation distortion, which is uncorrelated with the desired undistorted part {  by the symmetry of the Gaussian distribution}. In this case, treating this distortion as Gaussian noise with the same PSD in addition to the original additive noise leads to a lower bound on the information rate as discussed next. \\

Furthermore, when combining the signal from $N$ such direct detection antennas, with channel vector $\B{h}$ having  $\left\|\B{h}\right\|^2=N$, the desired signal part corresponding to $\rho/B$ combines coherently, while the noise and even the distortion part do not combine coherently,  since $\sum_{i=1}^N h_i^* |h_ix(t)|^2=\mathcal{O}(\sqrt{N})|x(t)^2| $ for most practical channels  (except for a planar array in the front-fire direction). Note that $\B{h}^{\rm H}$ is not the optimal linear receiver (see next Section) due to the spatially colored second order  distortion and the following results can be further improved. Assuming for simplicity $\B{h}^{\rm H}$ as combiner, we get the following lower bound on the achievable (proof technique similar to Theorem~\ref{LBtheorem} in the next section)
\begin{equation}
R_{\rm DD} \geq \int_{0}^{B} C_\varepsilon(f) {\rm d} f,
\label{RDD}
\end{equation}
with 
\begin{equation}
\begin{aligned}
&C_\varepsilon(f) =
\log_2 \!\left(\!1 \!+\! \frac{  N\frac{\rho}{B} }{N_0 + \frac{ \sigma^2 |G(f)|^2 \otimes |G(-f)|^2 }{\varepsilon P_{\rm LO} |G(\frac{B}{2}-|f|)|^2}  \left(N_0 +  \frac{\rho}{B}\right)^2 } \right) \!,
\label{cepsilon}
\end{aligned}
\end{equation}
{  where notably $G(f)$ from (\ref{approx_transfer}) also depends on $\sigma$}. 
Further, we calculate for $-B\leq f \leq B $

\begin{equation}
\begin{aligned}
&\displaystyle |G(f)|^2 \otimes |G(-f)|^2 = \\
&\int\limits_{\max (-\frac{B}{2},-\frac{B}{2}+f)}^{\min (\frac{B}{2},\frac{B}{2}+f)}  \!\!\!\!\!\!\!\!\!\!\!\! \frac{1}{ (2f'/B-1)^2 +  \sigma^{2}  }  \frac{1}{ (2(f'-f)/B-1)^2 +  \sigma^{2} } {\rm d}f' \\
&\!\!= \!\frac{B \arctan_\pi\!\!\left(  \frac{2(2-\bar{f}) (\sigma^{2}+ \bar{f})}{\sigma (\sigma^{2} -4 + \bar{f} (6-\bar{f}) )} \right) }{  2 \sigma (4\sigma^{2}  + \bar{f}^2 )  } \! + \! \frac{B \ln\!\! \left( \! \frac{\sigma^{2}(4+\sigma^{2})}{(\sigma^{2}+(\bar{f}-2)^2 )(\sigma^{2} +\bar{f}^2 )} \!\right) }{  2 \bar{f} (4\sigma^{2}  + \bar{f}^2 )  },
\end{aligned}
\end{equation} 
where $\bar{f}=2|f/B|$ and the $\arctan_\pi$ function is the arctangent mapping on the codomain $[0,\pi]$.

 Now, assuming the inverse square law for the LO coefficient $\varepsilon$, i.e. free space attenuation, $\varepsilon \propto N^{-2}$, then we state the following theorem  on the convergence of the achievable rate to the ideal case for large $N$.


\newtheorem {theorem1}{Theorem}
\begin {theorem1}
\label{dualitytheorem}
If $\sigma \leq o(\varepsilon)$ then the achievable rate $R_{\rm DD}$ with wireless LO distribution converges to $B \log_2(1+\frac{N \rho}{BN_0})$ when $\varepsilon$ converges to zero.  {  In other words, the resistive losses have to go to zero faster than the LO power portion $\varepsilon$ when the number of antennas increases.} 
\end {theorem1}
\begin{proof}
The expression { $\sigma^2\varepsilon^{-1} |G(f)|^2 \otimes |G(-f)|^2 \stackrel{f\neq 0}{=} \sigma^2\varepsilon^{-1}  \mathcal{O}(\sigma^{-1})=\mathcal{O}(\sigma \varepsilon^{-1})$}  present in the denominator of the logarithm in (\ref{cepsilon})  converges to $0$ for all frequencies $f\neq 0$ if $\sigma \leq o(\varepsilon)$. We deduce that
\begin{equation}
  \lim\limits_{\varepsilon \rightarrow 0, f \neq 0}   C_\varepsilon(f) = C= \log_2(1+\frac{N\rho}{BN_0}).
\end{equation}
Since  $C_\varepsilon(f)$ converges to $C$ almost everywhere except at $\{0\}$ which has zero Lebesgue measure, and  since $C_\varepsilon(f)$  is majorized by $C=\log_2(1+N\rho/(BN_0))$, we have by the dominated convergence theorem \cite{bartle95} the following lower bound for $R_{\rm DD}$

\begin{equation}
\begin{aligned}
\displaystyle \lim_{\varepsilon \rightarrow 0} \int_{0}^{B} C_\varepsilon(f)   {\rm d}f &= \displaystyle  \int_{0}^{B} \lim_{\varepsilon \rightarrow 0} C_\varepsilon(f)   {\rm d}f = B \log_2\left(1+\frac{N\rho}{BN_0}\right).
\end{aligned}
\end{equation}
As $B \log_2\left(1+\frac{N\rho}{BN_0}\right)$ is at the same time an upper bound on $R_{\rm DD}$ by the data processing theorem, the theorem is proved. 
\end{proof}

The reasoning used to prove the achievability of the ideal capacity with a wireless LO signal does not explicitly make use of the properties of the propagation channel. Therefore it can be easily generalized to more general channels with multiple users and frequency selectivity.  \\

\begin{figure}[tb]
\centering
\psfrag{N}[c][c]{$N$}
\hspace{-1mm}\includegraphics[width=0.43\textwidth]{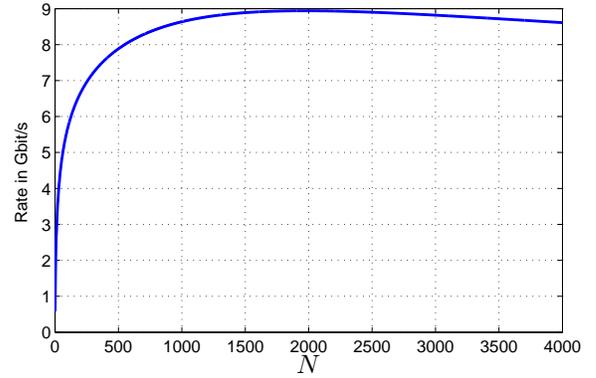}
\vspace{-2mm}
\caption{Achievable rates with wireless synchronization vs. $N$ for fixed bandpass filter with $\sigma=0.05$, $\rho/(BN_0)=0$dB, $B=1$GHz, $P_{\rm LO}=1\mu$W, $N_0=-174.2$dBm/Hz, Noise figure $=3$dB. \vspace{-0.5cm}}
\label{rate_vs_N}
\end{figure}

From a practical point of view the Q-factor $\sigma^{-1}$ of the filter $G(f)$ is limited and cannot be arbitrarily high. Therefore, we consider in Fig.~\ref{rate_vs_N} the achievable rate (\ref{RDD}) for fixed $\sigma=0.05$ and $P_{\rm LO}=1\mu$W as function of the number of antennas $N$, while the LO attenuation scales as $\varepsilon=N^{-2}$. We observe that the maximum is achieved at around $N=2000$, which is very encouraging in terms of the use of wireless synchronization for massive MIMO.

\section{System with One-bit ADCs}

In this section, we aim at further simplifying the RF front-end by using one-bit ADCs to convert the baseband signal into the digital domain. It is worth mentioning that sampling the RF signal directly at passband without frequency mixing  would also be possible with one-bit ADCs, but the analog filtering, the timing jitter and the required sample-and-hold circuit might be critical. Therefore, we focus in the following on coarse quantization of the baseband signal, while the wirelessly synchronized mixing can  potentially be used for the down-conversion.  For the analysis, we consider $K$ single antenna users  and $N$  receive antennas in the uplink. For simplicity, we assume a frequency  flat channel, even though the concept can be generalized to a frequency selective setting. \\

After propagation through the channel, and a desirably simple analog preprocessing, the unquantized received signal prior to the ADCs reads as
\begin{equation}
    \B{y} = \B{H} \cdot \B{x} +  \B{z},
\end{equation}
while the one-bit converted signal for the digital processing is
\begin{equation}
\B{r}=\frac{1}{\sqrt{2}}{\rm sign}({\rm Re} \{\B{y}\})+\frac{\rm j}{\sqrt{2}}{\rm sign}({\rm Im} \{\B{y}\}),
\end{equation}
where $\B{z} \in \mathbb{C}^N$ is the noise vector having i.i.d. elements with unit variance, $\B{H} =[\B{h}_1,\ldots,\B{h}_K] \in \mathbb{C}^{N \times K}$ comprises the user channels  $\B{h}_k$, $k=1, \ldots,K$, that are assumed to be known at the receiver and $\B{x}$ is the transmitted data which is assumed to be i.i.d. Gaussian distributed with variance $\rho$ (representing the SNR)\footnote{The Gaussian assumption is not essential here as the noise dominates the signal on each antenna especially in the mmwave bands. It is made to provide a lower bound on the performance and characterize the gap to the ideal case.}.  Similarly, the entries of $\B{H}$ are assumed to be i.i.d. Gaussian with unit variance \footnote{The asymptotic results derived later turn out to be  quite useful also for sparse channels according to our simulations.}.  


\section{Bussgang decomposition and performance characterization for the one-bit system}
\label{buss_decomp}
The Bussgang theorem \cite{Bussgang} implies that one can decompose the output of the nonlinear quantizer $\B{r}=Q(\B{y})$ into a
desired signal component and an uncorrelated distortion $\B{e}$
\begin{equation}
\begin{aligned} 
  \B{r}= \B{D} \B{y} + \B{e},
\end{aligned}
\end{equation}
where $ \B{D}$ can be obtained from the linear minimum mean square error (MMSE) estimation of $\B{r}$ from $ \B{y}$
\begin{equation}
\begin{aligned} 
   \B{D}={\rm E}[\B{r} \B{y}^{\rm H}]{\rm E}[\B{y} \B{y}^{\rm H}]^{-1}= \B{C}_{ry} \B{C}_{y}^{-1},
\end{aligned}
\end{equation}
and  the distortion error $\B{e}$ has the following correlation matrix 
\begin{equation}
\begin{aligned} 
  \B{C}_{e}&={\rm E}[(\B{r}- \B{D} \B{y})  (\B{r}- \B{D} \B{y})^{\rm H}] \\
            &= \B{C}_{r} - \B{C}_{ry}  \B{D}^{\rm H} -   \B{D} \B{C}_{yr} +  \B{D} \B{C}_{y}   \B{D}^{\rm H}  \\
            &= \B{C}_{r} - \B{C}_{ry}\B{C}_{y}^{-1}\B{C}_{yr}.
\end{aligned}
\label{C_e}
\end{equation}
The Bussgang decomposition ensures also that  $\B{e}$ and $\B{x}$ are uncorrelated if $\B{x}$ is jointly Gaussian with $\B{y}$. To prove that,  we use the fact, that $\B{e}=Q(\B{y})-\B{Dy}$ is a deterministic function of $\B{y}$, and thus, when conditioned on $\B{y}$, is independent of all other signals. That is
 \begin{equation}
 \begin{aligned}
 {\rm E}[\B{x}\cdot \B{e}^{\rm H} ] &=  {\rm E}_{y}[{\rm E}[\B{x}\cdot \B{e}^{\rm H}| \B{y}]]  \\
                                              &= {\rm E}_{y}[{\rm E}[\B{x} | \B{y}]\cdot {\rm E}[\B{e}^{\rm H}| \B{y}]]  \\
                                              &= {\rm E}_{y}[   \B{C}_{xy} \B{C}_{y}^{-1} \cdot {\rm E}[\B{e}^{\rm H}| \B{y}]] \\
                                              &= \B{C}_{xy} \B{C}_{y}^{-1} {\rm E}[\B{y} \cdot \B{e}^{\rm H}]  
                                              = \B{0},
 \end{aligned}
 \end{equation}
 where we used the fact that the Bayesian estimator ${\rm E}[\B{x} | \B{y}]$ corresponds to a linear estimator for jointly Gaussian signals $\B{x}$ and $\B{y}$.
Based on this decomposition, the channel output $\B{r}$ can be written as function of the channel input in the following form 
\begin{equation}
\begin{aligned} 
\B{r}&=  \B{D} \B{y} + \B{e} \\
     &=  \B{D} \B{H} \B{x} +  \B{D} \B{z}+ \B{e}  \\
     &=\B{H}' \B{x}  + \B{z}',
\end{aligned} 
\end{equation}
where we introduced the  effective channel
\begin{equation}
\begin{aligned}  
\B{H}' &=   \B{D} \B{H}  =  \B{C}_{ry} \B{C}_{y}^{-1}\B{H},
\end{aligned} 
\label{H_eff}
\end{equation}
and the non-Gaussian effective noise $\B{z}'$ with the covariance matrix
\begin{equation}
\begin{aligned} 
\B{C}_{z'} & = \B{C}_{e} +  \B{D}  \B{D}^{\rm H}    \\
&=  \B{C}_{r} - \B{C}_{ry}\B{C}_{y}^{-1}\B{C}_{yr} +   \B{C}_{ry} \B{C}_{y}^{-1}   \B{C}_{y}^{-1}\B{C}_{yr} \\
&= \B{C}_{r} - \rho \B{H}' \B{H}'^{\rm H}.
\end{aligned} 
\label{R_eff}
\end{equation}
 Based on this decomposition, the following lower bound has been derived in \cite{mezghani2007,mezghani_2012_isit}. 
\newtheorem {theorem2}[theorem1]{Theorem}
\begin {theorem2}
\label{LBtheorem}
For the quantized system with i.i.d. Gaussian input $\B{x}$ of covariance matrix $\B{C}_x=\rho{\bf I}$,  we have 
\begin{equation}
\begin{aligned} 
I(x_k;\B{r}) \geq R_k,
\end{aligned} 
\label{c_lo_b}
\end{equation}
with 
\begin{equation}
\begin{aligned} 
R_k= \log_2 (1  + \rho \B{h}_k'^{\rm H} (\B{C}_r- \rho \B{h}_k'\B{h}_k'^{\rm H})^{-1} \B{h}_k' )= \log_2 (1  + \gamma_k),
\end{aligned} 
\label{cap_g}
\end{equation}
wehre $\B{H}'$ and $\B{C}_{z'}$ are given in (\ref{H_eff}) and (\ref{R_eff}) respectively. 
\end{theorem2}
\begin{proof}
We first introduce the linear minimum mean square error  (LMMSE) estimate of $\B{x}$ given the quantized observation $\B{r}$ reading as
\begin{equation}
\begin{aligned}
 \hat{\B{x}}&= \B{C}_{xr}\B{C}_{r}^{-1}\B{r} = \B{C}_{x}\B{H}'^{\rm H} (\B{H}' \B{C}_{x} \B{H}'^{\rm H} + \B{C}_{z'})^{-1}  \B{r}. 
\end{aligned}
\end{equation}
Then we have the lower on the achievable rate per user
\begin{eqnarray}
I(x_k,\B{r})&=&h(x_k)-h(\B{x}|\B{r})=   h(x_k)-h(x_k|\B{r}) \nonumber\\
&=&  h(x_k)-h(x_k-\hat{x_k}|\B{r})   \label{x_hat_det} \\  
&\geq&  h(x_k)-h(\underbrace{x_k-\hat{x}_k}_{\epsilon_k}) \label{conditioning} \\
&\geq& \log_2\frac{c_{x_k}}{c_{\epsilon_k}} \label{conditioning2}.
\end{eqnarray}
We get (\ref{x_hat_det}) as $\hat{\B{x}}$ is a deterministic function of $\B{r}$. Since conditioning reduces entropy, we obtain inequality (\ref{conditioning}). On the other hand, The second term in (\ref{conditioning}) is upper bounded by the entropy of a Gaussian random variable whose covariance is equal to the error variance $c_{\epsilon_k} = \rho - \rho^2 \B{h}_k'^{\rm H} \B{C}_r^{-1} \B{h}_k'$ of the linear minimum mean square error (MMSE) estimate of $x_k$, leading to (\ref{conditioning2}). Finally we obtain the lower bound on the mutual information as in (\ref{cap_g}) using the matrix inversion lemma.
\end{proof}
\subsection{Bussgang decomposition based on the Price's theorem and Taylor expansion}
An elegant way to perform the Bussgang decomposition for general types of nonlinearities is to use Price's theorem \cite{price} which can provide the derivatives of the output covariance matrix $\B{C}_r$ as function of the covariance matrix $\B{C}_y$
as follows \cite{Bos_96} 

\begin{equation}
 \frac{\partial^k c_{r_ir_j}}{\partial c_{y_iy_j}^k}={\rm E}\left[ \frac{\partial^{k} Q(y_i)}{\partial y_i^k}\frac{\partial^{k} Q(y_j)^*}{\partial y_j^{k,*}}\right].
\end{equation}
 Since the derivative of the quantization operation is described by the Dirac-delta function, the calculation  of the first order derivative or higher of $[\B{C}_r]_{i,j}$ with respect to $[\B{C}_y]_{i,j}$ with $i \neq j$ is possible in closed form. Therefore, even if calculating $\B{C}_r$ in closed form might be not possible, one can still determine the following Taylor expansion around ${\rm nondiag}(\B{C}_y)=\B{0}$
\begin{equation}
 {\rm nondiag}(\B{C}_r)\!= \!\underbrace{{\rm nondiag}(\B{\Sigma}_1  \! \circ  \! \B{C}_y)}_{\textrm{desired part}} \!+\! \underbrace{{\rm nondiag}(\B{\Sigma}_3 \! \circ  \!\B{C}_y^{\circ 3}) \!+\! \cdots}_{\textrm{distortion uncorrelated with }\B{x}},
 \label{taylor}
\end{equation} 
where $\B{\Sigma}_\ell$ are matrices that are only a function of ${\rm diag}(\B{C}_y)$. Due to the odd symmetry of the quantization function,  the even-order terms vanish. The expansion is decomposed into two part as explained in the following. 
 Bussgang's theorem states that the matrix $\B{C}_{ry}$ is row-wise proportional to $\B{C}_{y}$. Consequently, the desired (undistorted) part $\B{C}_{ry} \B{C}_{y}^{-1} \B{C}_{yr}$  in the decomposition (\ref{C_e}) is a diagonally scaled version of $\B{C}_{y}$ which turns to be given by (again based on the Price's theorem)
\begin{equation}
 [\B{C}_{ry} \B{C}_{y}^{-1} \B{C}_{yr}]_{i,j}=  \Big. \frac{\partial c_{r_ir_j}}{\partial c_{y_iy_j}} \Big|_{c_{y_iy_j}=0} \cdot c_{y_iy_j}= [\B{\Sigma}_1]_{i,j} \cdot c_{y_iy_j}
\end{equation} 
for $i \neq j$. In other words, the Bussgang decomposition (\ref{C_e}) can  be interpreted as extracting  the linear term in the Taylor expansion (\ref{taylor}) and considering it as the desired signal part, while treating the remaining higher order terms as additive distortion noise. To study the impact of the nonlinearity, it is very useful to consider the first and third order terms. These terms will be studied in the next two subsections.     

In the special case of a one-bit symmetric quantizer, it is even possible to express $\B{C}_r$ in closed form. In fact, due to the  classical \emph{arcsine law} \cite{price}, the output of a decision device $r_{i,{\rm Re/Im}}=\frac{1}{\sqrt{2}}{\rm sign}[y_{i,{\rm Re/Im}}]\in \{-\frac{1}{\sqrt{2}},\frac{1}{\sqrt{2}}\}$ applied to a multi-variable Gaussian input $\B{y}$ has the following correlation matrix 
\begin{equation}
\B{C}_{r}=\frac{2}{\pi} \left[ \arcsin\left( \textrm{diag}(\B{C}_{y})^{-\frac{1}{2}}\B{C}_{y}\textrm{diag}(\B{C}_{y})^{-\frac{1}{2}}\right)\right],
\end{equation}
with $\B{C}_{y}= \rho  \B{H} \B{H}^{\rm H}+  {\bf I}$,
where the arcsine function is applied element-wise to its matrix argument.
Additionally, the correlation matrix between the input and the output of the 1-bit quantizer can be obtained as \cite{Bussgang}  
\begin{equation}
\B{C}_{ry}=\sqrt{\frac{2}{\pi}} \textrm{diag}(\B{C}_{y})^{-\frac{1}{2}} \B{C}_{y}.
\end{equation}
Then, we get the effective channel from (\ref{H_eff}) as
\begin{equation}
\B{H}'=  \sqrt{\frac{2}{\pi}} \textrm{diag}(\B{C}_{y})^{-\frac{1}{2}} \B{H},
\label{H_prime}
\end{equation}
while the effective noise covariance in (\ref{R_eff}) becomes
\begin{equation}
\begin{aligned}
\B{C}_{z'}=&\frac{2}{\pi} \left[ \arcsin\left( \textrm{diag}(\B{C}_{y})^{-\frac{1}{2}}\B{C}_{y}\textrm{diag}(\B{C}_{y})^{-\frac{1}{2}}\right)\right]- \\
& \frac{2}{\pi}  \rho \textrm{diag}(\B{C}_{y})^{-\frac{1}{2}} \B{H} \B{H}^{\rm H} \textrm{diag}(\B{C}_{y})^{-\frac{1}{2}} .
\end{aligned}
\label{Cz_exact}
\end{equation}
The \emph{arcsine law} is however not tractable when characterizing the performance in the large system limit. Therefore, we resort to the first order and third order approximations introduced previously. 
\subsection{First order approximation}
For low SNR per antenna, which is relevant for mmwave applications, we can make the first order approximation following (\ref{taylor}) 
\begin{equation}
\B{C}_{r} \stackrel{\rho \ll 1}{\approx}  \textrm{diag}(\B{C}_{y})^{-\frac{1}{2}} \left[ \frac{2}{\pi}\B{C}_{y}+(1-\frac{2}{\pi})   \textrm{diag}(\B{C}_{y})\right]  \textrm{diag}(\B{C}_{y})^{-\frac{1}{2}},
\end{equation}
leading to the uncorrelated effective noise  from (\ref{R_eff})
\begin{equation}
\B{C}_{z'}\! \stackrel{\rho \ll 1}{\approx} \! \textrm{diag}(\B{C}_{y})^{-\frac{1}{2}} \!\left[ \! \frac{2}{\pi} {\bf I}+(1\!-\!\frac{2}{\pi})   \textrm{diag}(\B{C}_{y})\right]  \textrm{diag}(\B{C}_{y})^{-\frac{1}{2}} \!.
\end{equation}
This approximation is also only valid when $N$ is not significantly larger than $K$, an observation that will be discussed later on. 
Next, using the fact that $\rho \ll 1$, we obtain the effective  signal-to-interference-noise-and-distortion ratio  (SINDR)
\begin{align}
&\gamma_k = \rho \B{h}_k'^{\rm H}\left(\B{C}_{z'} + \rho  \sum\limits_{k'\neq k} \B{h}_{k'}'\B{h}_{k'}'^{\rm H} \right)^{-1} \B{h}_k' \label{gamma_exact} \\
&\stackrel{\rho \ll 1}{\approx} \rho \B{h}_k^{\rm H} \Bigg( {\bf I}+(\frac{\pi}{2}\!-\!1)  {\rm diag} ({\bf I} \!+ \!\rho \B{H}\B{H}^{\rm H})\! + \!\rho  \!\sum\limits_{k'\neq k}\! \B{h}_{k'}\B{h}_{k'}^{\rm H} \Bigg)^{-1} \!\!\B{h}_k.
\end{align} 
Further, following the common massive MIMO assumption $N \gg K \gg 1 $, we have ${\rm diag}( \B{H}\B{H}^{\rm H}) \rightarrow K {\bf I} $ and $\B{h}_{k'}^{\rm H}\B{h}_{k} \rightarrow 0$ for $k' \neq k$, and $N$ for $k' = k$, which yields the asymptotic first order result
\begin{equation}
 \gamma \approx {N\frac{\rho}{1+K\rho}}\left({\frac{\pi}{2}}-\frac{K\rho}{1+K\rho}\right)^{-1}= \frac{N \rho}{\frac{\pi}{2}+(\frac{\pi}{2}-1)K\rho},
 \label{Cz_approx}
\end{equation}
 corresponding to the well known performance loss of factor $2/\pi$ ($\approx\!-1.96$dB) of the one-bit system compared to the ideal case at low SNR per antenna $\rho$ \cite{mezghaniisit2007,Yongzhiuplink,mezghani_2012_isit}. In the next subsections, we provide a more accurate third order approximation and discuss the validity of the first order approximation. In addition, we investigate the efficiency of linear processing treating the quantization error as additive noise. In fact, the third order approximation provides an indication of whether or not linear processing is appropriate for quantized massive MIMO, and when advanced DSP is required to maintain the large antenna gain.
\subsection{Third order approximation}
From (\ref{gamma_exact}), (\ref{Cz_exact}), (\ref{H_prime}), we obtain the following expression using the third order Taylor expansion of the arcsine function (c.f. (\ref{taylor})), together with the large system limit ${\rm diag}( \B{C}_y) \rightarrow (1+ K\rho) {\bf I} $:
\begin{equation}
\begin{aligned}
&\gamma_k \!\!\stackrel{\rho \ll 1}{\approx} \!\!\frac{\rho}{1+K\rho} \B{h}_k^{\rm H}  \Bigg(  (\frac{\pi}{2}-\frac{K\rho}{1+K\rho}) {\bf I}  +  \Bigg.  \\
 &\Bigg.   \!\!\frac{1}{6}\! \left(\!\frac{\rho}{1+K\rho}\right)^{\!3}\!{\rm nondiag}((\B{H}\B{H}^{\rm H})^{\circ 3}) \!+\! \frac{\rho}{1+K\rho} \! \sum\limits_{k'\neq k}\! \B{h}_{k'}\B{h}_{k'}^{\rm H} \!\Bigg)^{\!\!-1} \!\! \B{h}_k.
 \end{aligned}
 \label{third_order1}
\end{equation}
Next, we consider the matrix  ${\rm nondiag}((\B{H}\B{H}^{\rm H})^{\circ 3})$ and aim at simplifying it by identifying the significant distortion part in the direction of the channel $\B{h}_{k}$
\begin{equation}
\begin{aligned}
&\!\!\!\!\!\!\!\!\!\!\!\!\B{h}_{k}^{\rm H}{\rm nondiag}((\B{H}\B{H}^{\rm H})^{\circ 3}) \B{h}_{k}  \\
&\stackrel{\rm (a)}{\approx} 3\B{h}_{k}^{\rm H}{\rm nondiag}((\B{H}\B{H}^{\rm H})^{\circ 2}\circ  \B{h}_{k}\B{h}_{k}^{\rm H} ) \B{h}_{k}  \\
&\stackrel{\rm (b)}{\approx}  3\B{h}_{k}^{\rm H}{\rm nondiag}({\rm E}[(\B{H}\B{H}^{\rm H})^{\circ 2}]\circ  \B{h}_{k}\B{h}_{k}^{\rm H} ) \B{h}_{k}  \\
&=3\B{h}_{k}^{\rm H}{\rm nondiag}((\frac{K}{2}+\frac{{\rm j}K}{2})\circ  \B{h}_{k}\B{h}_{k}^{\rm H} ) \B{h}_{k}  \\
&=\frac{3K}{2}\B{h}_{k}^{\rm H}{\rm nondiag}( \B{h}_{k}\B{h}_{k}^{\rm H} ) \B{h}_{k},  
 \end{aligned}
\end{equation}
where, in (a), we neglect the terms that do not combine coherently with respect to the direction $\B{h}_k$, and in (b), we approximate the matrix $(\B{H}\B{H}^{\rm H})^{\circ 2}$ by its expectation. Thus, we can replace ${\rm nondiag}((\B{H}\B{H}^{\rm H})^{\circ 3})$ by $\frac{3K}{2}{\rm nondiag}( \B{h}_{k}\B{h}_{k}^{\rm H} ) \approx \frac{3K}{2} \B{h}_{k}\B{h}_{k}^{\rm H}$ (approximation in the Frobenius norm sense)   in  (\ref{third_order1}). Further, we neglect the inter-user interference in  (\ref{third_order1}) via the assumption $N \gg K$ to obtain the final expression
\begin{equation}
\begin{aligned}
 \gamma \approx &N {\frac{\rho}{1+K\rho}} \left({ \frac{\pi}{2}}-\frac{K\rho}{1+K\rho}\right)^{-1} \cdot  \\ 
&\left( 1-\frac{1.5N\cdot K}{6 \left({ \frac{\pi}{2}}-\frac{K\rho}{1+K\rho}\right)\left(\frac{1+K\rho}{\rho}\right)^{\!\! 3} +1.5 N\cdot K} \right).
 \end{aligned}
 \label{third_order}
\end{equation}
This formula reveals that the processing gain is actually bounded, as some distortion terms combine coherently in the user's channel direction when $N \rightarrow \infty$:
\begin{equation}
\begin{aligned}
\lim\limits_{N\rightarrow \infty} \gamma \approx \frac{4}{K} \left(\frac{1+K\rho}{\rho}\right)^2.
\end{aligned}
\end{equation}
 The linear behavior predicted by the first order approximation (\ref{Cz_approx}) is only valid up to a certain $N$  and holds longer the smaller $\rho$ is. The approach presented here can be also  applied to higher resolution or other type of nonlinearities.
\begin{figure}[h]
\psfrag{N}[c][c]{$N$}
\psfrag{antenna gain}[c][c]{$\gamma/\rho$}
\centerline{\includegraphics[width=8cm]{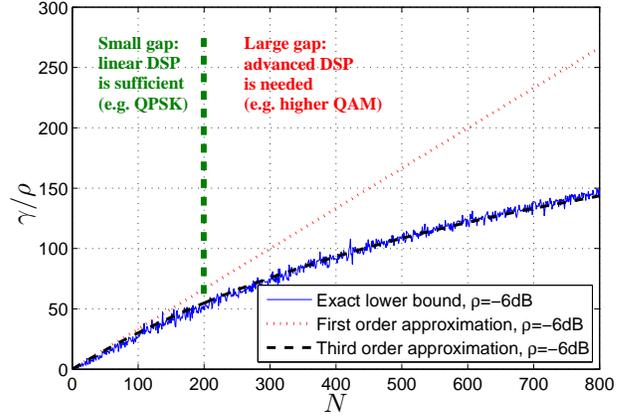}}
\caption{Achievable linear processing gain (\ref{gamma_exact}) versus $N$ for i.i.d. channel and $\rho=-6$dB using the exact formula (\ref{gamma_exact}) based on (\ref{Cz_exact}) and the approximations (\ref{Cz_approx}) and (\ref{third_order}). For QPSK, linear processing can be sufficient; for 16QAM, however, nonlinear advanced processing \cite{mezghani2008,Hong_2017,juncil2015near,Wen_Wu_2015} is required.}
\label{fig_G1}
\end{figure}
\begin{figure}[h]
\psfrag{N}[c][c]{$N$}
\psfrag{gamma per stream}[c][c]{$\gamma/\rho$}
\centerline{\includegraphics[width=8cm]{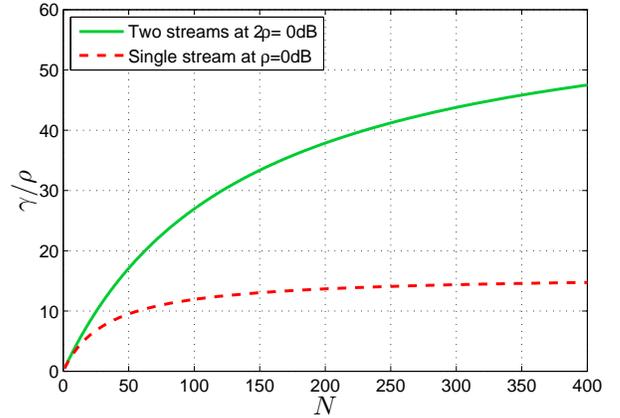}}
\caption{Achievable processing gain versus $N$ for $\rho=0$dB (one stream, $K=1$) and  $\rho=-3$dB (two streams/dual polarization,  $K=2$). \vspace{-1cm}}
\label{fig_G2}
\end{figure}
\subsection{Validity of the uncorrelated distortion assumption and the benefits of spatial multi-streaming}

Let us first consider some numerical examples. Then, we will discuss the validity of the first order approximation (\ref{Cz_approx}) that assumes uncorrelated distortion error and draw key consequences for the design of such low resolution systems. In fact, the validity of this approximation is also an indicator of whether or not linear DSP would be sufficient. The achievable linear processing gain $\gamma/\rho$ from (\ref{gamma_exact}) versus $N$ for $K=10$ and $\rho=-6$dB is plotted in Fig.~\ref{fig_G1} using the exact formula (\ref{gamma_exact}), the first order approximation (\ref{Cz_approx}) and the first order approximation (\ref{third_order}). 
The third order formula seems to reflect the behavior of $\gamma$ accurately as we observe in Fig.~\ref{fig_G1}, and it is also useful for evaluating the performance of linear detection methods for one-bit massive MIMO systems with $N$ antennas and $\rho$ as the SNR per data stream and per antenna.   
In contrast,  we observe an increasing gap between the first order approximation (\ref{Cz_approx}) and the exact formula (\ref{gamma_exact}), since the correlations of the quantization errors become more effective with more antennas compared to the noise. The observed increasing gap suggests that with a very large number of antennas, the nonlinear effects do not completely vanish and one should consider the use of advanced nonlinear DSP as developed in \cite{mezghani2008,Hong_2017,juncil2015near,Wen_Wu_2015} to further maintain the array gain, particularly if higher order modulation is desired which requires higher processing gain.  This is in contrast to the common assumption that linear processing is nearly optimal with larger $N$ in the ideal case. This assumption is not necessarily sufficient for low resolution receivers even for a single user scenario if higher order modulation schemes are intended. 
By comparing (\ref{Cz_approx}) and (\ref{third_order}), we deduce the following proposition

\newtheorem {proposition1}{Proposition}
\begin {proposition1}
In addition to the common massive MIMO assumption, $N \gg K$, the following condition is required for the near-optimality of linear processing with one-bit ADCs:
\begin{equation}
 N_{\rm linear~DSP} \gg { \frac{\pi}{4}} \sqrt{K \gamma^3}= { \frac{\pi}{4}} \sqrt{K (2^R-1)^3}.
\end{equation} 
\end {proposition1}
\begin{proof} 
Linear processing is nearly optimal when the approximation (\ref{Cz_approx}) is valid, i.e., when the i.i.d. quantization noise assumption holds. To ensure this, we deduce from the more accurate approximation (\ref{third_order}) the following condition:
\begin{equation}
\begin{aligned}
&1.5 N\cdot K \ll \\
& 6 \left({ \frac{\pi}{2}}-\frac{K\rho}{1+K\rho}\right)\left(\frac{1+K\rho}{\rho}\right)^{\!\! 3} \approx 6 \left({ \frac{\pi}{2}}-\frac{K\rho}{1+K\rho}\right)^{-2} \frac{N^3}{\gamma^3}.
\end{aligned}
\end{equation}
After straightforward simplifications, we obtain the result.
\end{proof}

The required number of antennas increases cubically with the desired $\gamma$, which might become inconvenient (1000s of antennas) for higher-order modulation schemes. To rely on linear DSP, this suggests that the user's terminals should instead aim at reducing their initial SNR per dimension, by using the entire bandwidth and the time interval and having more streams instead of using higher order modulation and/or concentrating the signals in space, time or frequency. In fact, making use of all available dimensions decorrelates the quantization error, which is extremely beneficial as we can observe in Fig.~\ref{fig_G2}, where very surprisingly  $\gamma$ after processing can be higher with two parallel streams than with one stream for the same total power and a larger number of antennas. In other words, the achievable rate is more than doubled when going from one-stream to double-stream transmission and linear DSP becomes more efficient. This effect is essential in low resolution receivers and can be even more impactful with more superimposed independent signals in space, time or frequency.  It is worth noting that two spatial streams are even possible in a line-of-sight condition based on dual polarization. 



\section{Conclusion}
We considered the use of wireless LO synchronization and one-bit receivers for reducing the RF complexity of large array base stations in the uplink. We showed that  wireless LO synchronization is possible with limited losses for more than 1000 antennas even with practical band-pass filters and strict emission requirements.  Low resolution receivers with a large number of antennas are intended to operate at low to moderate SNR per antenna and information dimension. To this end, it is very desirable that users simultaneously exploit the entire available bandwidth and spatial dimensions. That implies that the popular FDMA or TDMA schemes are not appropriate for low resolution systems while SDMA and spatial multi-streaming strategies are extremely beneficial. In fact,  the use of higher order modulations, which is principally possible with coarsely quantized massive antenna arrays, generally requires advanced nonlinear processing techniques, while  multi-streaming with QPSK modulation can still be performed with linear techniques at low RF and DSP complexity. 
\bibliographystyle{IEEEtran}     
\bibliography{references}{}

\end{document}